\def\aap{\mbox{\bf{Astron.\ Astrophys.}}}

\def\apjl{\mbox{\bf{Astrophys.\ J.\ L.}}}

\def\apjs{\mbox{\bf{Astrophys.\ J.\ Suppl.}}}

\documentclass[traditabstract]{aa}

\usepackage{amsmath}
\usepackage{longtable, lscape}
\usepackage{graphicx}
\usepackage{txfonts}
\usepackage{natbib}
\usepackage{color}

\bibpunct{(}{)}{;}{a}{}{,} 

\newcommand{\Msun}{\mbox{$~\mathrm{M}_{\odot}$}}

\newcommand{\asem}{\mbox{$\alpha_{\rm SEM}$}}

\newcommand{\fmu}{\mbox{$f_{\mu}$}}
\newcommand{\fc}{\mbox{$f_{c}$}}

\begin{document}

\title{ Rotational mixing in massive binaries:} 
\subtitle{detached short-period systems}
   \author{ S.~E. de Mink\inst{1}, M. Cantiello\inst{1},
   N. Langer\inst{1,2}, O.~R. Pols\inst{1}, I. Brott\inst{1},
   S.-Ch. Yoon\inst{3}}


 \institute{Astronomical Institute, Utrecht University, PO Box 80000,
   3508 TA Utrecht, The Netherlands \and Argelander-Institut
   f\"{u}r Astronomie der Universit\"{a}t Bonn, Auf dem H\"{u}gel 71,
   53121 Bonn, Germany \and Dep. of Astronomy \& Astrophysics,
   Univ. of California, Santa Cruz, CA95064, USA~\\
   S.E.deMink@uu.nl, M.Cantiello@uu.nl, N.Langer@uu.nl,
     O.R.pols@uu.nl, I.Brott@uu.nl, scyoon@ucolick.org }
 \authorrunning{ S.~E. de Mink, M. Cantiello, N. Langer, et al.}
 \date{Received : 28/11/2008 ; accepted : 02/02/2009 }

  \abstract{ 
     Models of rotating single stars can successfully account for
    a wide variety of observed stellar phenomena, such as the surface
    enhancements of N and He observed in massive main-sequence
    stars. However, recent observations have questioned the idea that
    rotational mixing is the main process responsible for the surface
    enhancements, emphasizing the need for a strong and conclusive
    test for rotational mixing.

    We investigate the consequences of rotational mixing for massive
    main-sequence stars in short-period binaries. In these systems the
    tides are thought to spin up the stars to rapid rotation,
    synchronous with their orbital revolution.
    We use a state-of-the-art stellar evolution code including the
    effect of rotational mixing, tides, and magnetic fields. We adopt
    a rotational mixing efficiency that has been calibrated against
    observations of rotating stars under the assumption that
    rotational mixing is the main process responsible for the observed
    surface abundances.

    We find that the primaries of massive close binaries ($M_1 \approx
    20\Msun$, $P_{\rm orb}\lesssim$ 3 days) are expected to show
    significant enhancements in nitrogen (up to 0.6 dex in the Small
    Magellanic Cloud) for a significant fraction of their core
    hydrogen-burning lifetime.  
    We propose using such systems to test the concept of rotational
    mixing. As these short-period binaries often show eclipses, their
    parameters can be determined with high accuracy.

    For the primary stars of more massive and very close systems ($M_1
    \approx 50\Msun$, $P_{\rm orb}\lesssim$ 2 days) we find that
    centrally produced helium is efficiently mixed throughout the
    envelope. The star remains blue and compact during the main
    sequence evolution and stays within its Roche lobe. It is the less
    massive star, in which the effects of rotational mixing are less
    pronounced, which fills its Roche lobe first, contrary to what
    standard binary evolution theory predicts. The primaries will
    appear as ``Wolf-Rayet stars in disguise'': core hydrogen-burning
    stars with strongly enhanced He and N at the surface.  We propose
    that this evolution path provides an alternative channel for the
    formation of tight Wolf-Rayet binaries with a main-sequence
    companion and might explain massive black hole binaries such as
    the intriguing system M33~X-7.  }
   \maketitle

\section{Introduction}\label{sec:intro}

The rotation rate is considered as one of the main initial stellar
parameters, along with mass and metallicity, which determine the
fate of single stars.
Rotation deforms the star to an oblate shape \citep[of which Achernar
is an extreme example, see][]{Domiciano+03}, it interplays with the
mass loss from the star \citep[e.g.][]{Friend+86, Langer98,
Maeder+00} and it induces instabilities in the interior leading to
turbulent mixing in otherwise stable layers \citep[e.g.][]{Maeder+97}.
As rotationally induced mixing can bring processed material from the
core to the surface, it has been proposed as explanation for observed
surface abundance anomalies, such as a nitrogen enrichment found in
several massive main-sequence stars
\citep[e.g.][]{Walborn76,Maeder+Meynet+review00,Heger+Langer00}.
In models of rapidly rotating massive stars, rotational mixing can
efficiently mix the centrally produced helium throughout the stellar
envelope. Instead of expanding during core H burning as non-rotating
models do, they stay compact, become more luminous and move blue-wards
in the Hertzsprung-Russell diagram.  This type of evolution is
commonly referred to as (quasi-)chemically homogeneous evolution
\citep{Maeder87, Yoon+Langer05}. Some over-luminous O and
WNh\footnote{Wolf-Rayet stars with evidence of enhanced nitrogen and
hydrogen in their spectra \citep[e.g.][]{schnurr+08}} in the
Magellanic Clouds in the Magellanic Clouds have been put forward as
examples pf this type of evolution
\citep{Bouret+03,Walborn+04,Mokiem+07,Martins+08}. This
alternative evolutionary scenario has been proposed as a way to create
rapidly rotating massive helium stars as the possible progenitors of
long gamma-ray bursts within the collapsar scenario
\citep{Yoon+Langer05, Woosley+06, Cantiello+07}.

Multiple attempts have been made to constrain the efficiency of
rotational mixing \citep[e.g.][]{Gies+Lambert92, Fliegner+96,
Daflon+01, Venn+02, Korn+02, Huang+Gies06,Mendell+06}, but often these
attempts remained inconclusive due to limited sample sizes or a strong
bias towards stars with low projected velocities.
The VLT-FLAMES survey provided, for the first time, a large sample of
massive stars with accurate abundance determinations, covering a wide
range of projected rotational velocities
\citep{Evans+05,Hunter+08rotandN}. This sample was used to calibrate
the efficiency of rotational mixing, under the assumption that
rotational mixing is the main process responsible for the observed
enhancements \citep{Brott+09}. 

These authors performed a population synthesis, based on detailed
models of rotating single stars, to reproduce the properties of the
VLT-FLAMES sample.  They find that their models cannot account for a
large number (20\% of the sample) of highly nitrogen-enriched, slow
rotators and a group of evolved fast rotators which are relatively
non-enriched (a further 20\% of the sample). This raises the question
whether other processes play an important role in explaining the
nitrogen enhancements of massive main-sequence stars, such as mass
transfer in binaries \citep{Langer+08}.

Regardless of the successes of rotating stellar models to explain a
variety of stellar phenomena \citep{Maeder+Meynet+review00},
rotational mixing is still a matter of debate. Clearly a conclusive
observational test for the concept of rotational mixing is needed.  In
this paper we propose to use eclipsing binaries for this purpose.


Eclipsing binaries have frequently been used to test stellar evolution
models as they provide accurate stellar masses, radii and effective
temperatures.  Even beyond our own Galaxy, in the Magellanic Clouds,
stellar parameters of O and early B stars have been determined with
accuracies of 10\% \citep{Harries+03,Hilditch+05}, which have been
used to test binary evolution models \citep[e.g.][]{demink+07}.  As
rotational mixing is more important in more massive stars
\citep[e.g.][]{Heger+00}, it is a major advantage to know the stellar
masses for quantitative testing of the efficiency of rotational
mixing. 

In close binaries with orbital periods $P_{\rm orbit}$ less than a
few days, the tides are so strong that the stars rotate synchronously
with the orbital motion: $P_{\rm spin} = P_{\rm orbit}$. With 
stellar radii known from eclipse measurements, this enables us to
determine the rotational velocity directly from the orbital period. This is
the second important advantage of using binaries for testing rotational
mixing with respect to single stars.  For single stars fitting of
spectral lines allows only for the determination of $\varv \sin i$, where
$\varv$ is the rotational velocity at the equator and $i$ the inclination
of the rotation axis, which is generally not known.

Here, we propose to use eclipsing binaries, consisting of two  detached
main-sequence stars. Detailed calculations of binary evolution show
that if one of the stars fills its Roche lobe during the main
sequence, it does not detach again before hydrogen is exhausted in the
core, except maybe for a very short thermal timescale
\citep{wellstein+01, demink+07}. Turning this around we find that, in
a binary with two detached main-sequence stars, we can safely exclude
the occurrence of previous mass transfer.  In other words, the fact
that a system consists of two detached main-sequence stars, constrains
the evolutionary history. In contrast, for a fast rotating apparently
single star we do not know whether the star was born as a rapidly
rotating single star, or whether the rapid rotation is the result of
mass transfer in a binary or of a binary merger.  The companion, if
still present, may be very hard to detect, being a faint low-mass star
in a wide orbit%
\footnote{Abundance determinations of boron, if available, may be used
  to distinguish between binary effects and pure rotational mixing,
  see \citet{Fliegner+96}}.

If the spectra of a binary are of high quality, one can determine the
surface abundances of the two components \citep[e.g.][]{Leushin88N,
  Pavlovski+Hensberge05, Rauw+05}.  These surface abundances, together
with accurate determinations of the stellar parameters and the orbital
period have the potential of strongly constraining the efficiency of
rotational mixing.

The evolution of close massive binaries has been modeled by various
groups \citep[e.g.][and references
therein]{Podsiadlowski+92,Pols+94,wellstein+01,Wellstein+Langer99,
  Nelson+Eggleton01,Belczynski+02,Petrovic+05_WR,vanbeveren+07,demink+07}.
In this work we use a detailed binary evolution code to predict the
surface abundances for massive detached close binaries.  In addition
we discuss models of very massive close binaries, in which rotational
mixing can be so efficient that the change in chemical profile leads
to changes in the stellar structure.

\section{Stellar evolution code}\label{sec:code}  

We model the evolution of rotating massive stars using the 1D
hydrodynamic stellar evolution code described by \citet{Yoon+06} and
\citet{Petrovic+05_GRB}, which includes the effects of rotation on the
stellar structure and the transport of angular momentum and chemical
species via rotationally induced hydrodynamic instabilities
\citep{Heger+00}.
The rotational instabilities considered are: dynamical shear, secular
shear, Eddington-Sweet circulation and the Goldreich-Schubert-Fricke
instability \citep{Heger+00}. Two processes dominate rotational mixing
in massive stars: Eddington-Sweet circulation, large-scale meridional
currents resulting from the thermal imbalance between pole and equator
characteristic of rotating stars \citep{Vonzeipel24, Eddington25,
Eddington26, Vogt25} and shear mixing, eddies, that can form between
two layers of the star rotating at different angular velocity
\citep[e.g.][]{Zahn74}.
 We take into account angular momentum transport by magnetic
torques as proposed by \citet{Spruit02} as these can successfully
provide the coupling between the stellar core and envelope necessary to
explain the observed spins in young compact stellar remnants
\citep{Heger+05, Petrovic+05_GRB, Suijs+08}.  

 Mixing of chemical species and the transport of angular momentum are
implemented as diffusion processes. For convection we assume a mixing
length parameter $\alpha_{\rm MLT}= 1.5$. In semi-convective regions
we assume efficient mixing \citep[$\asem=1.0$, as defined
in][]{Langer91}. The turbulent viscosity is determined as the sum of
the convective and semi-convective diffusion coefficients and those
that arise from the rotationally induced instabilities.  The
inhibiting effect of gradients in the mean molecular weight on
rotational mixing is decreased by a factor $\fmu=0.1$ \citep{Yoon+06}.

\citet{Brott+09} calibrated the efficiency of rotationally induced mixing 
using data from the VLT-FLAMES survey of massive stars, assuming that
rotational mixing is the main process responsible for the observed
enhancements. To reproduce the extension of the main sequence in the
Hertzsprung-Russell diagram, they assumed overshooting of 0.355 times
the pressure scale height.  The main parameter responsible for the
efficiency of rotational mixing is
\fc, defined as the contribution of rotationally induced instabilities
to the diffusion of chemical species \citep{Heger+00}. \citet{Brott+09}
found that $\fc=0.0228$ is needed to reproduce the spread in N
abundances observed in the VLT-FLAMES data
\citep{Hunter+08rotandN}.  We do not consider possible mixing by
 instabilities related to magnetic buoyancy due to winding up of the
 magnetic field lines through differential rotation \citep{Spruit02}
 as this leads to too efficient mixing \citep{Hunter+08rotandN}.

Metallicity-dependent mass loss in the form of stellar winds has been
included as in \citet{Yoon+06, Brott+09}. For the associated angular
momentum loss we assume that mass is lost with the specific angular
momentum equal to the latitudinally averaged specific angular
momentum of the surface layer. The mass loss is thus assumed to be
independent of latitude.

The effect of mass and angular momentum loss on the binary orbit is 
computed according to \citet{Podsiadlowski+92}, with the specific
angular momentum of the wind calculated according to
\citet{Brookshaw+93}. We assume that the orbit is circular and that
the spins of the stars are perpendicular to the orbital plane.
  Tidal interaction is modeled as described in
\citet{Detmers+08} using the timescale for synchronization by
turbulent viscosity \citep{Zahn77}, see also
Section~\ref{sec:tides}.
Angular momentum is transported between the surface layers and the
inner regions by magnetic torques and rotational instabilities. At the
onset of central H burning, we assume that the rotation of the stars
is rigid and synchronized with the orbital motion.

\subsection*{Initial Composition}

We discuss single stellar models starting with three different initial
compositions, representing the composition of the Small and Large
Magellanic Cloud (abbreviated as SMC and LMC) and the Galaxy (GAL)
following the approach by \citet{Brott+09} to which we refer for
details. For C, O, Mg and Si we use the average abundances determined
for stars in the VLT-FLAMES survey \citep{Hunter+07}. For the
Magellanic Clouds we adopt the N abundance measured for HII regions
which is in agreement with the lowest observed N abundances in the
VLT-FLAMES sample \citep[and references therein]{Hunter+07}. For the
remaining heavy elements we adopt solar abundances by
\citet{Asplund+05} scaled down by 0.7 dex for the SMC and 0.4 dex
for the LMC. For our binary models we have assumed the SMC composition.
The opacity tables of OPAL \citep{Iglesias+Rogers96} are adopted, where
the Fe abundance is used to interpolate between the tables of
different metallicity.

\begin{figure}[t]
\resizebox{\hsize}{!}{ \includegraphics[angle=90]{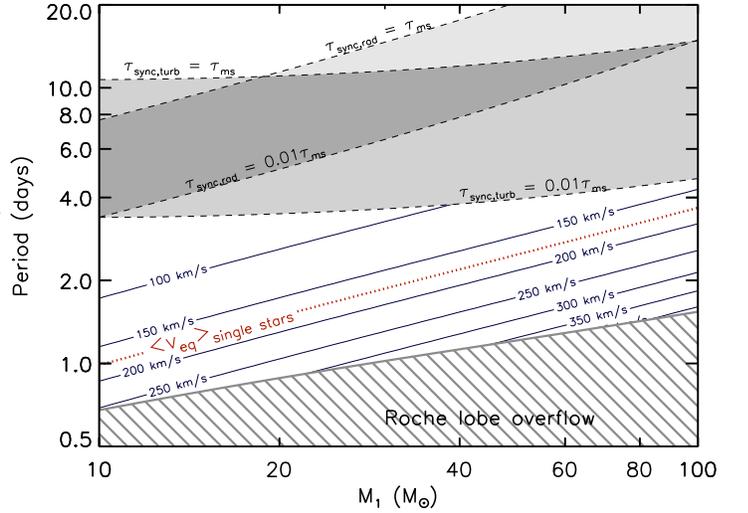}}
\caption{ \label{tides} Timescales for tidal synchronization.  The
  four dashed lines indicate where the timescales for tidal
  synchronization by turbulent viscosity $\tau_{\rm sync,turb}$ and
  radiative dissipation $ \tau_{\rm sync,rad}$ for zero-age
  main-sequence stars are equal to (one percent of) the main-sequence
  lifetime $\tau_{\rm ms}$ of the most massive star of a binary
  system, see Sect.~\ref{sec:tides} for details.  In the region below
  the gray-shaded bands, tides quickly synchronize the stellar
  rotation with the orbit. The resulting equatorial velocity in
  kms$^{-1}$ for the primary star is indicated with contour levels,
  assuming the radius at the onset of hydrogen burning.  We have
  assumed a metallicity of Z=0.004 and a mass ratio ${\rm M}_2/{\rm
    M}_1=0.75$, to be consistent with the models presented in
  Sect.~\ref{sec:sin}.  The average velocity measured for apparently
  single stars in the SMC, $\langle \varv_{\rm eq}\rangle =
  175$kms$^{-1}$, is plotted with a dotted line.  Very short orbital
  periods are excluded (hashed region) as the stars fill their Roche
  lobe at zero-age. }
\end{figure}

\section{Stellar rotation under the influence of tides}\label{sec:tides}

In a close binary, angular momentum and kinetic energy can be exchanged
between the two stars and their orbit through tides.  The system tends to
evolve towards a state of minimum mechanical energy due to dissipative
processes.  This is an equilibrium state, where the orbit is circular,
the spins of the stars are aligned and perpendicular to the orbital
plane and the stars are in synchronous rotation with the orbital motion,
such that $P_{\rm spin} = P_{\rm orbit}$.  How quickly this
equilibrium state is approached depends on the efficiency of the processes
responsible for the energy dissipation.

The most efficient form of energy dissipation takes place in turbulent
regions of the star, such as convective layers, where the kinetic
energy of the large-scale flow induced by the tides cascades down to
smaller and smaller scales, until it is dissipated into heat.
\citet{Zahn77,Zahn89} estimated the timescale for synchronization due
to this ``turbulent viscosity'' as a function of the ratio $q$ of the
mass of the companion star to the mass of the star under consideration
and on the ratio of the stellar radius $R$ over the separation $a$
between the two stars:
\begin{equation}
 \tau_{\rm sync,turb} = f_{\rm turb} \, q^{-2}\, \left({R \over a}\right)^{-6} \text{ year.}
\end{equation} 
The proportionality factor $f_{\rm turb}$ depends on the structure of
the star, on the location of the turbulent layers and on the timescale
for dissipation of kinetic energy, which is uncertain. Nevertheless,
the dependence of the time scale on $R/a$ is so steep that,
approximating $f_{\rm turb} \approx 1$, \citet{Zahn77} showed that
this expression adequately explains the observed orbital period below
which tides lead to synchronization.

In the absence of turbulent viscosity, another dissipative process is
required in order to have efficient tides. If the star does not rotate
synchronously, it experiences a varying gravitational potential, which
triggers a large range of oscillations in the star.  These
oscillations are damped near the stellar surface by radiative
dissipation. \cite{Zahn75} derived the corresponding timescale for
synchronization:
\begin{equation}
 \tau_{\rm sync,rad} = f_{\rm rad} \,q^{-2} \,(1+q)^{-5/6} \,\left({R
 \over a}\right)^{-17/2} \text{ year,}
\end{equation}
\[
\text{where } f_{\rm rad}= 52^{-5/3} \,E_2 \,\left({I\over MR^2}\right)\, \left(GM\over R^3\right)^{-1/2}.
\]
Here, $M$ denotes the mass of the star under consideration, $I$ its
moment of inertia and $E_2$ the tidal coefficient, which is sensitive
to the structure of the star, in particular to the size of the
convective core: $E_2 \propto (R_{\rm core} / R )^{8}$
\citep{Zahn77}. 

The early-type massive stars considered in this work have radiative
envelopes \citep[except for convective zones just below the surface
which contain almost no mass, e.g.][]{Cantiello+09} and
synchronization by radiative damping has been proposed to be the most
efficient dissipation mechanism.  However, in fast rotating stars
turbulence is induced by rotational instabilities and one may argue
that the first timescale applies also to these stars
\citep{Toledano+07}.  In addition \citet{Witte+Savonije99} show
 that in some cases resonance locking can contribute to even more
 efficient tidal dissipation.

 In Figure~\ref{tides} we compare both synchronization timescales with
 the main-sequence lifetime of the more massive star in a binary.  For
 this plot we have used the radii of non-rotating zero-age main
 sequence stars assuming a metallicity of Z = 0.004 and a mass ratio
 of ${\rm M}_2/{\rm M}_1 = 0.75$. For $E_2$ we use an analytic
 approximation by \citet{Hurley+02}, for the Roche-lobe radius we use
 the analytic expression by \citet{Eggleton83}. The diagram shows that
 in binaries with orbital periods shorter than approximately 3-5 days,
 the timescale for synchronization is less than one percent of the
 main-sequence lifetime $\tau_{\rm MS}$, regardless of the actual
 process responsible for the synchronization. The resulting equatorial
 velocity of the primary star, assuming synchronous rotation, is given
 in kms$^{-1}$ along contour lines.

For comparison: the average equatorial velocity for apparently single
stars in the VLT-FLAMES survey with masses between 7 and 25\Msun~is
150~kms$^{-1}$ for the LMC and 175~kms$^{-1}$ for the SMC
\citep{Hunter+08rotandN}.  In most binaries tides will slow down the
rotation of the star, but in the tightest binaries the rotation rate
is higher than that for average single stars. Note that the binary
systems of interest in this work, for which the components have high
enough equatorial velocities to show significant surface abundance
changes by rotational mixing ($\gtrsim$100-150~kms$^{-1}$, see
Sect.~\ref{sec:sin}), are in the region where tides are very
efficient, according to both prescriptions for the synchronization
timescale.

\begin{figure*}
\centering
\includegraphics[width=0.9\textwidth]{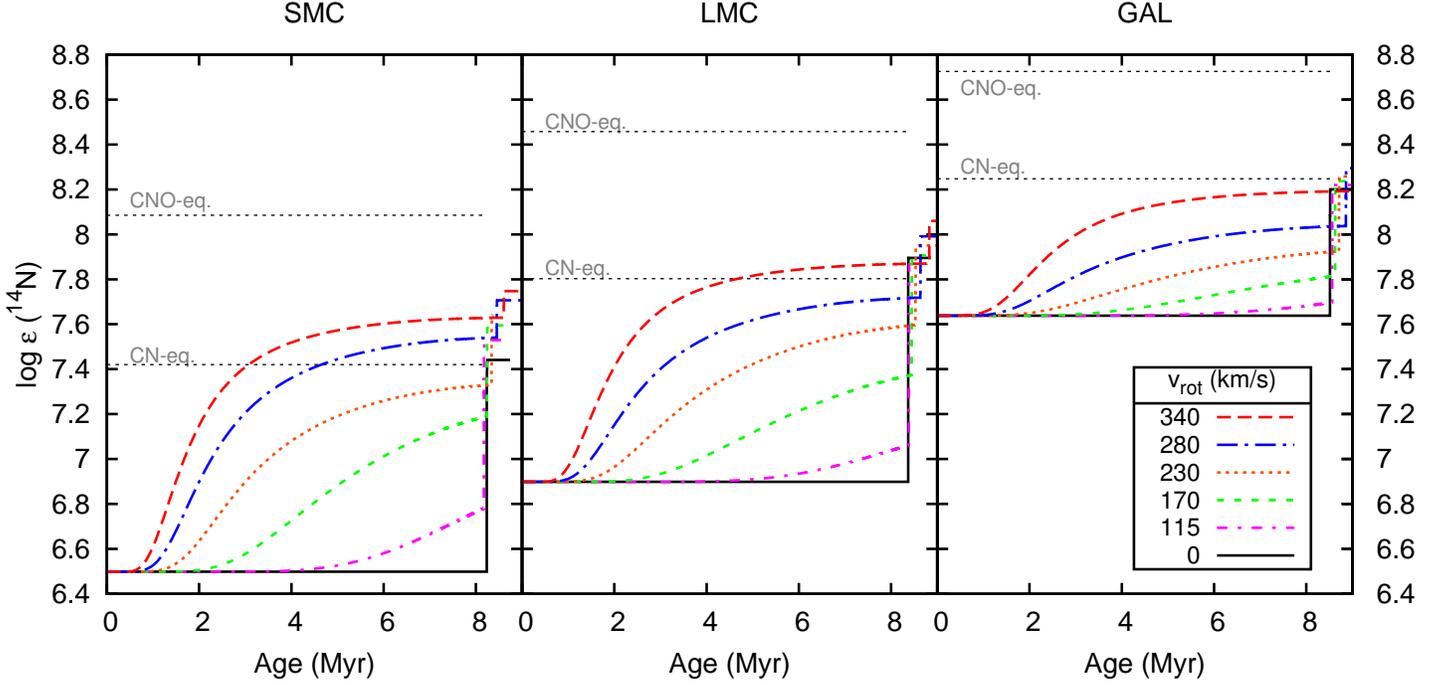}
  \caption{ Surface nitrogen abundance versus time for single stellar
  models of 20\Msun~with different (approximate) initial equatorial
  rotational velocities starting with the initial composition of the
  SMC (left panel), LMC (center panel) and GAL (right panel).  The
  nitrogen abundance is given in the conventional units: the logarithm
  of the number fraction of nitrogen $n_{\rm N}$ with respect to
  hydrogen $n_{\rm H}$, where the hydrogen abundance is fixed at
  $10^{12}$, i.e.  $\log \epsilon ({\rm N}) = \log_{10}[n_{\rm N} /
  n_{\rm H}] + 12$. \label{fig:mix_1}}
\end{figure*}

\section {Rotational mixing in single star models} 
\label{sec:models:sin}\label{sec:models}
\label{sec:sin} Any element produced or destroyed in the hot interior
of the star can in principle be used as a tracer of rotational
mixing. The amount by which the surface abundance of a particular
element changes depends on the location in the star where the element is
processed.

Helium is the most important element synthesized during core hydrogen
burning. By the time a substantial amount of helium has been formed in
the stellar core, a gradient in mean molecular weight has been
established at the interface between the core and the envelope, which
has an inhibiting effect on rotationally induced mixing. Models of
moderately fast rotating stars (with an equatorial velocity of about
170 kms$^{-1}$ and a mass of about 20\Msun) show that helium will only
appear at the surface towards the very end of the main-sequence
evolution. In more massive stars, the helium surface abundance may be
significantly enhanced but this is partly due to the strong stellar
winds. This makes helium not very suitable for testing the effect of
rotational mixing.

A better tracer for rotational mixing during the main-sequence
evolution is nitrogen, which is produced in the hot interior layers on
a very short time scale when carbon is converted into nitrogen by
CN-cycling. On a longer timescale (about 1-2 Myr in the center of a
20\Msun~star) the CNO cycle comes into equilibrium, leading to
additional N at the expense of both carbon and oxygen. Rotational
mixing can bring nitrogen to the surface, resulting in a gradual
increase of the nitrogen surface abundance over the main-sequence
lifetime (Fig.~\ref{fig:mix_1}).
Nitrogen can be measured in rotating B-type stars in the Magellanic
Clouds with accuracies of 0.2-0.25 dex \citep{Hunter+08rotandN}.
Carbon decreases accordingly, but cannot be measured as accurately and
is therefore less suitable as a tracer of rotational mixing.  Other
elements such as boron can be used, but these are less abundant and
may be hard to detect, especially in metal-poor environments such as
the Magellanic Clouds.

Rotational mixing becomes more efficient in higher-mass stars since
radiation pressure becomes more important, which helps to overcome the
entropy barrier at the interface between the core and the
envelope. Also, the ratio of the timescale for meridional circulation
with respect to the main-sequence lifetime decreases with increasing
mass \citep{Yoon+06}.  The disadvantage of using very massive stars as
test cases for rotational mixing is the uncertainty in their mass-loss
rates. As mass loss exposes deeper layers of the star, enriched in N
and He, it has qualitatively the same effect on the surface
composition as rotational mixing.  At lower metallicity mass loss in
the form of a radiatively driven stellar wind is reduced. Therefore,
also the uncertainty in the mass-loss rate is less important.

Figure~\ref{fig:mix_1} shows the nitrogen surface abundance as a
function of time for rotating single stellar models of 20\Msun~with
three different initial compositions, representative for the
relatively metal-poor composition of the Small and Large Magellanic
Cloud, and for a mixture representing stars in the Galaxy. The
equilibrium abundances for N in the center for CN- and CNO-cycling are
plotted as horizontal lines. The Galactic model with an equatorial
velocity of 230 kms$^{-1}$ shows nitrogen enhancements up to about 0.3
dex, while in the SMC model the N abundance increases by up to
approximately 0.8 dex.  The high C/N ratio in the Magellanic clouds is
partly responsible for this effect as it leads to a strong increase of
N during CN-cycling \citep{Brott+09}. Indeed, in the VLT-FLAMES
survey, \citet{Hunter+08rotandN} find a larger spread in N abundances
for the stars in the Magellanic Clouds than for ones in the Galactic
stars.  We conclude that the Magellanic Clouds are the most promising
location for testing rotational mixing, as the effect of rotational
mixing is most pronounced in the N surface abundances.

\begin{table*}  
  \caption{ Key properties of the massive binary evolution models
    (20\Msun$+$15\Msun) as defined and discussed in
    Sect.~\ref{bin:hil}.  \label{tab:smc} } \centering
\begin{tabular}{c c c c c c c c c c c c}         
\hline\hline                       
$P_{\rm orb}$ & 
$t_{\rm RL}$ & 
$t_{\rm delay}$ &  
$ t_{\rm RL} - t_{\rm enh} $  & 
$R/R_{\rm RL} $ &  
X$_{\rm He,center}$ & 
$ \log \epsilon ({\rm N}^{14})$ & 
$ ^{}{\rm B}_{\rm RL}/^{}{\rm B}_{\rm init}^\mathrm{a}$ & 
$ ^{}{\rm C}_{\rm RL}/^{}{\rm C}_{\rm init}^\mathrm{a}$ &
$ ^{}{\rm N}_{\rm RL}/^{}{\rm N}_{\rm init}^\mathrm{a}$ &
$\langle \varv_{\rm eq} \rangle^\mathrm{b}$ & 
$\varv_{\rm eq,RL}^\mathrm{b}$ \\
(d) & (Myr) & (Myr) &(Myr) & ($t=t_{\rm enh} $)   & ($t=t_{\rm RL} $) & ($t=t_{\rm RL} $) & & & &  (kms$^{-1}$) & (kms$^{-1}$)  \\
\hline 
  1.1 &   3.3 &   1.3 &   1.0 &   0.92&   0.42 &   7.12 &   0.11  &   0.87 &   4.17 &    243 &  267 \\
  1.2 &   3.9 &   1.6 &   1.2 &   0.90&   0.47 &   7.07 &   0.08  &   0.85 &   3.71 &    228 &  260 \\
  1.4 &   5.0 &   2.1 &   1.5 &   0.86&   0.55 &   6.98 &   0.06  &   0.82 &   3.04 &    203 &  248 \\
  1.6 &   5.5 &   2.7 &   1.0 &   0.86&   0.61 &   6.89 &   0.07  &   0.83 &   2.45 &    181 &  234 \\
  1.8 &   6.0 &   3.2 &   0.9 &   0.86&   0.65 &   6.85 &   0.07  &   0.83 &   2.27 &    166 &  228 \\
  2.0 &   6.3 &   3.7 &   0.6 &   0.87&   0.69 &   6.81 &   0.09  &   0.86 &   2.05 &    152 &  216 \\
  2.2 &   6.7 &   4.1 &   0.5 &   0.89&   0.73 &   6.79 &   0.10  &   0.87 &   1.96 &    141 &  212 \\
  2.4 &   6.9 &   4.5 &   0.4 &   0.90&   0.75 &   6.77 &   0.11  &   0.88 &   1.87 &    131 &  207 \\
  2.6 &   7.1 &   4.9 &   0.2 &   0.93&   0.78 &   6.74 &   0.13  &   0.90 &   1.74 &    123 &  199 \\
  2.8 &   7.2 &   5.2 &   0.1 &   0.96&   0.79 &   6.73 &   0.14  &   0.91 &   1.69 &    116 &  196 \\
  3.0 &   7.3 &   5.6 &   0.0 &   0.99&   0.81 &   6.70 &   0.15  &   0.92 &   1.59 &    110 &  190 \\
\hline 
\end{tabular}
\begin{list}{}{}
\item[$^{\mathrm{b}}$] The surface mass fraction at the onset of
    Roche-lobe overflow divided by the initial mass fraction of
    resp. boron ($^{10}$B + $^{11}$B), carbon ($^{12}$C) and nitrogen
    ($^{14}$N).
\item[$^{\mathrm{b}}$] The equatorial rotational velocity for the primary star,
    averaged over time from the onset of hydrogen burning until the
    start of mass transfer (one before last column) and at the onset
    of Roche-lobe overflow (last column).
\end{list}
\end{table*}

\section{Binary models}\label{sec:models:bin} \label {sec:bin}

In this section we present binary evolution models calculated with the
same set of input physics as the single stellar models presented in
Section~\ref{sec:sin}, taking into account the effects of mass and
angular momentum loss on the orbit and spin-orbit coupling by
tides. We assume a composition representative of the Small Magellanic
Cloud, which is relatively metal-poor and has a high carbon to
nitrogen ratio.  We discuss two specific model sets: systems with a
massive primary, $M_1=20\Msun$, in Sect.~\ref{bin:hil} and systems
with a very massive primary, $M_1=50\Msun$, in Sect.~\ref{bin:lmc}.

\begin{figure*}
\centering
 \includegraphics[width=8.5cm]{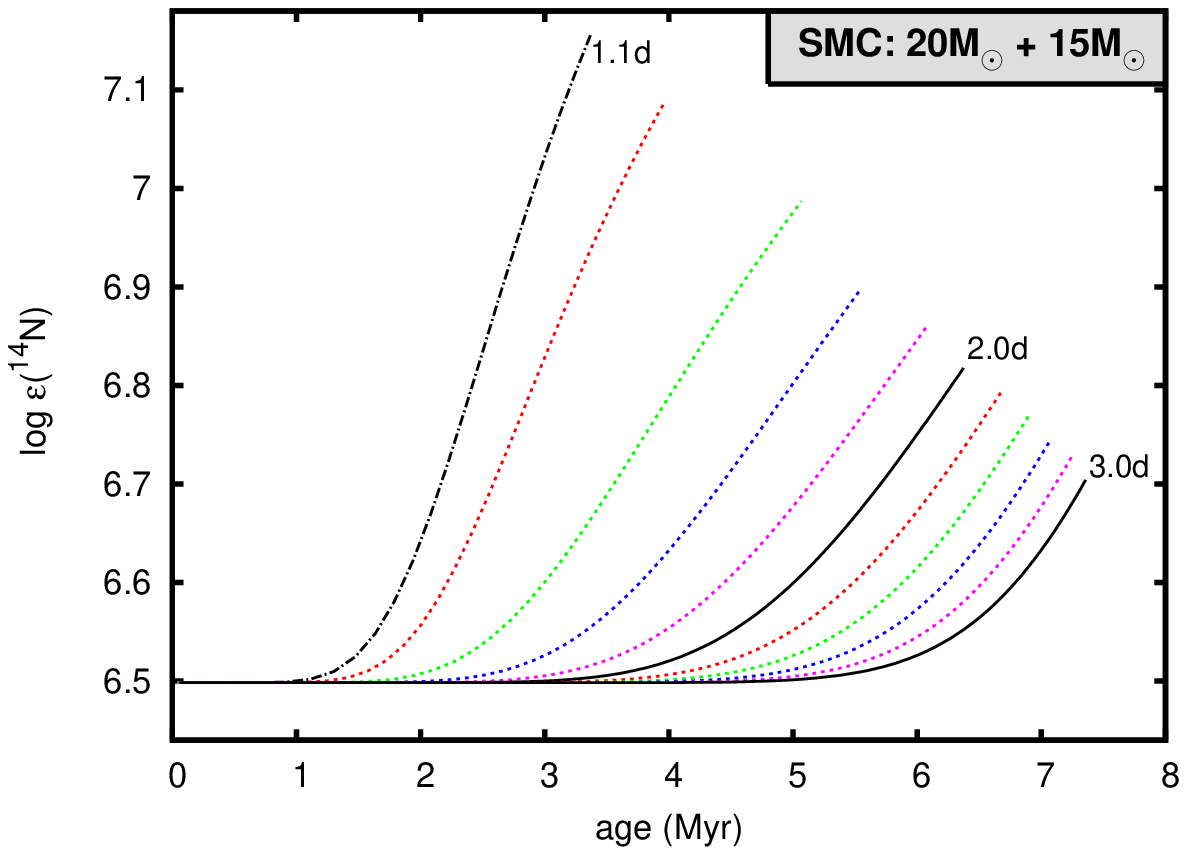}\includegraphics[width=8.5cm]{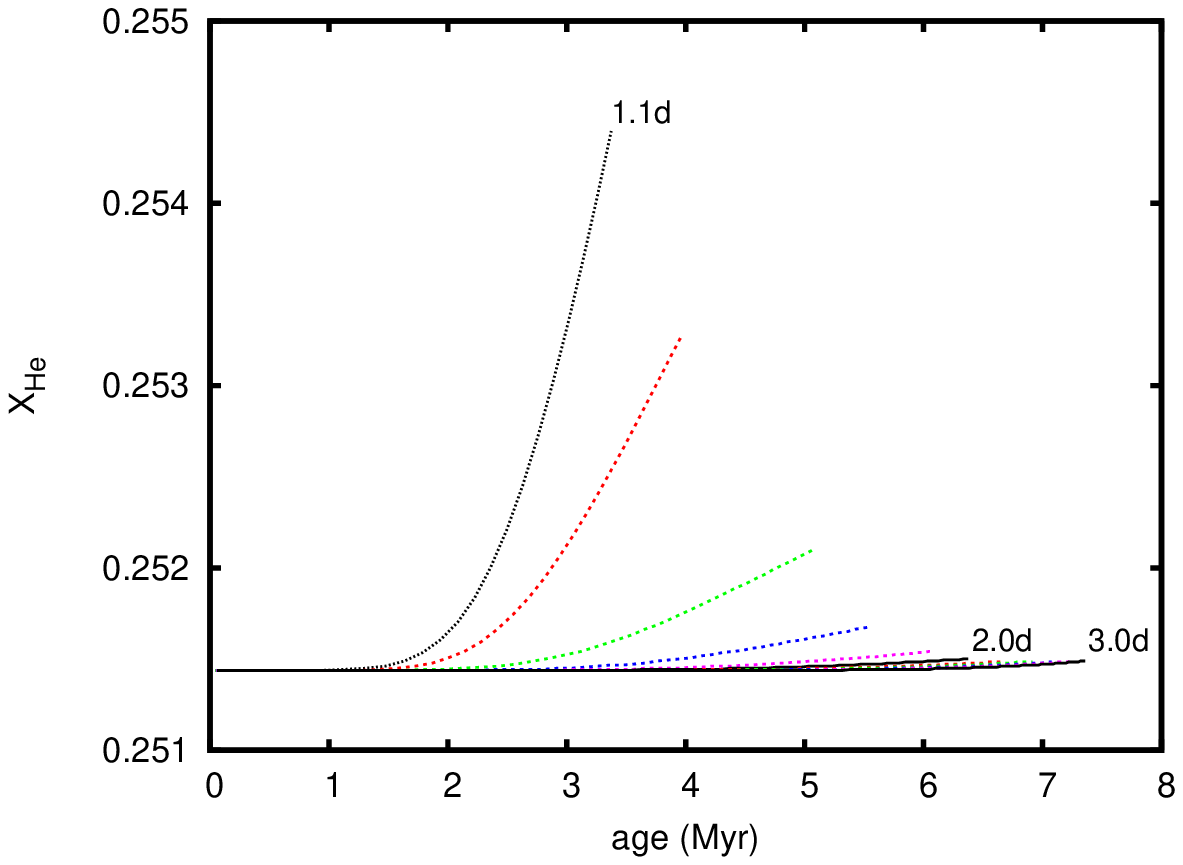} 
\includegraphics[width=8.5cm]{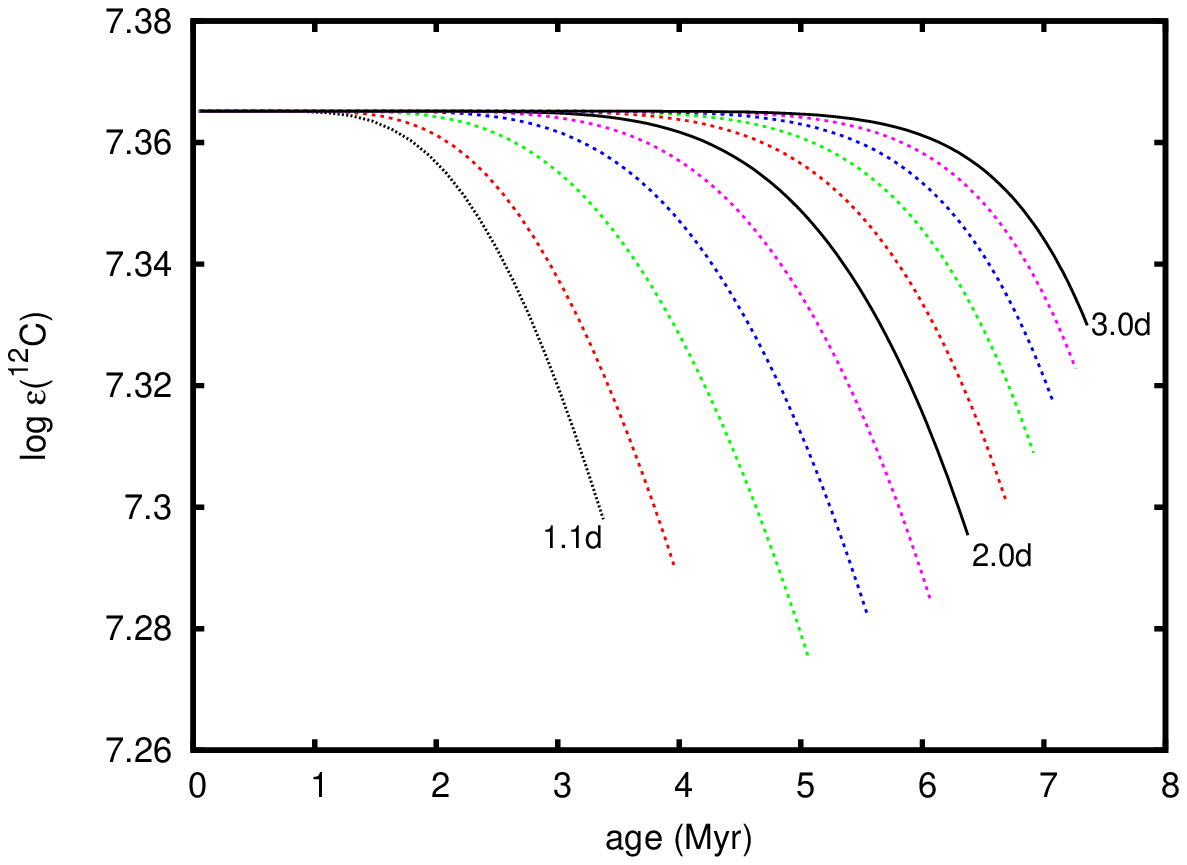}\includegraphics[width=8.5cm]{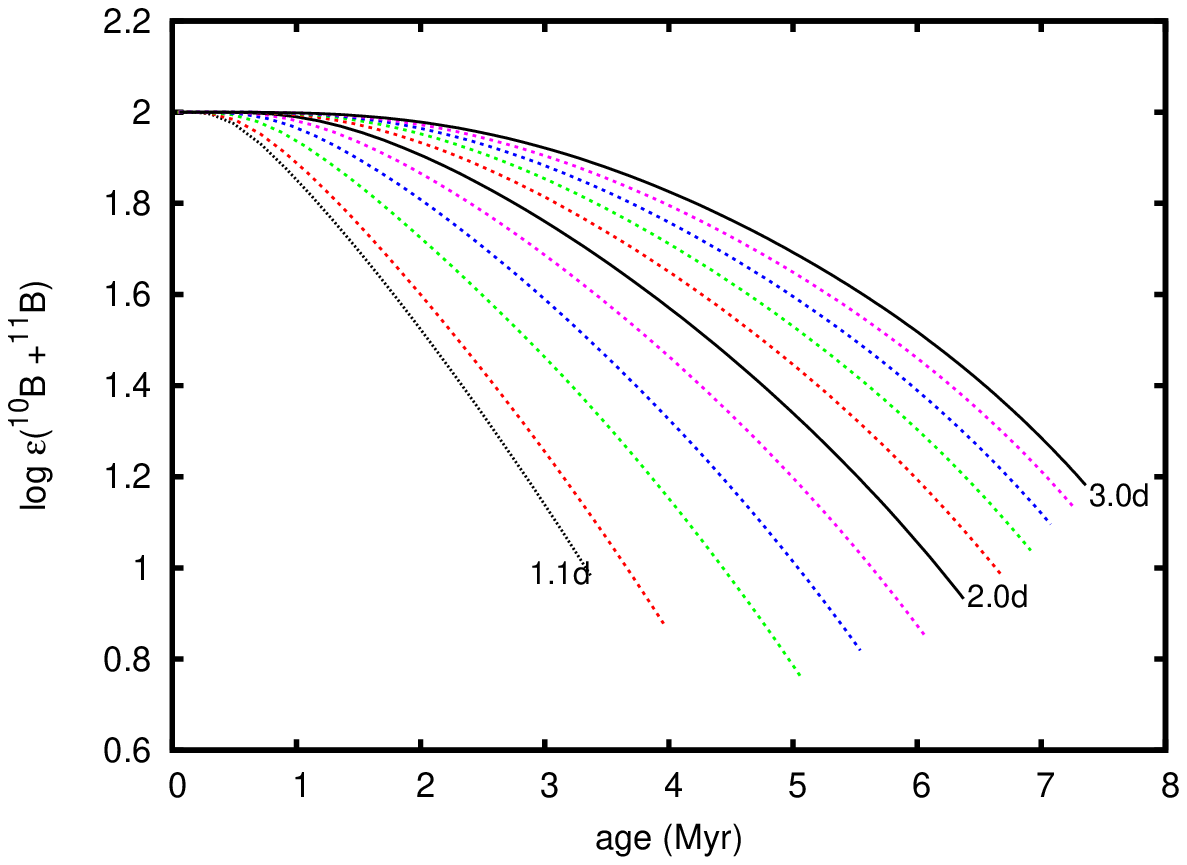}
\caption{Surface abundances of nitrogen ($^{14}$N),
  carbon ($^{12}$C), boron ($^{10}$B$ + ^{11}$B) and the mass fraction
  of helium at the surface versus time for a 20\Msun~star with a
  15\Msun~close companion. Note the different vertical scales. The
  abundance of an element X is given relative to hydrogen in the
  conventional units: $ \log \epsilon ({\rm X}) = \log_{10} (n_{\rm X}
  / n_{\rm H}) + 12$, where $n_X$ and $n_H$ refer to the number
  fractions.  The different lines show the evolution assuming initial
  orbital periods  of 1.1 days and between 1.2 and 3 days, with a
  spacing of 0.2 days.  The tracks are plotted from the onset of
  central H burning until the onset of Roche-lobe overflow.  See
  also Table~\ref{tab:smc}.  \label{fig:abun}}
\end{figure*}

\subsection { Massive binaries (20\Msun$+$15\Msun) }\label {bin:hil}\label {sec:hil}
With the Small Magellanic Cloud sample of double-lined eclipsing
binaries by  \citet{Harries+03} and \citet{Hilditch+05} in mind, which
contains 21 detached systems%
     \footnote{Possibly only 20 systems are detached.  For two of the
       systems an alternative semi-detached solution exists.  For one
       of these systems a comparison to binary evolution models
       including the effects of mass transfer showed that the
       semi-detached solution was more consistent than the detached
       solution \citep{demink+07}.}
     with orbital periods ranging from 1--4 days and primary masses
     ranging from 7--23\Msun, we chose to model the following binary
     systems. For the mass of the primary component we adopt 20\Msun,
     for the secondary component 15\Msun.  The evolution of both stars
     is followed starting at the onset of central hydrogen burning, at
     $t = 0$, until the primary star fills its Roche lobe, at $t =
     t_{\rm RL}$.  We adopt orbital periods up to 3 days. In these
     systems the tides are efficient enough to keep both stars in
     synchronous rotation with the orbit (Sect.~\ref{sec:tides}).

     \subsubsection*{Surface abundances} 
	
     The abundances of nitrogen, carbon, helium and boron at the
     surface of the primary star are shown as a function of time in
     Fig.~\ref{fig:abun}, see also Tab.~\ref{tab:smc}.  The nitrogen
     abundances at the surface starts to increase after about 1~Myr
     for the 1.1~day binary, and after about 5~Myr for the 3.0~day
     binary.  This time delay $t_{\rm delay}$ is the time it takes to
     transport the nitrogen from the deeper layers, where it is
     produced, to the surface\footnote{We define $t_{\rm delay}$ as
     the age at which the surface nitrogen abundance is enhanced by
     0.01 dex.}. The largest enhancement, 0.6 dex, is achieved in the
     1.1~day binary.  The shorter the orbital period, the faster the
     rotation of the stars, the more efficient rotational mixing and
     the faster the surface abundances change with time.  On the other
     hand in the systems with short orbital periods the stars fill
     their Roche lobe at an earlier stage, leaving less time to modify
     their surface abundances.

     The typical uncertainty in the nitrogen abundance determinations
     for stars in the VLT-FLAMES survey is about 0.2 dex. Therefore,
     to be able to detect whether nitrogen is enhanced, it should
     exceed the initial nitrogen abundance by 0.2 dex.  The age at
     which this lower limit for detecting a surface nitrogen
     enhancement is exceeded is denoted as $t_{\rm enh}$ in
     Table~\ref{tab:smc}. Systems with orbital periods shorter than 3
     days all reach surface enhancements of 0.2 dex, at the latest
     just before they fill their Roche lobe. The time span during
     which the primary can be observed with a surface nitrogen
     abundance exceeding 6.7 dex, listed as $ t_{\rm RL} - t_{\rm
     enh}$ in Table~\ref{tab:smc}, is highest for the 1.4 day system,
     1.5 Myr.  For detached tidally locked main-sequence binaries with
     initial orbital periods greater than about 3 days we do not expect
     detectable N enhancements on the basis of these models: their
     rotation rates and therefore the efficiency of rotational mixing
     is too low.

     While the nitrogen abundance increases, the carbon abundance
     decreases accordingly, see Fig.~\ref{fig:abun}. It decreases by
     less than 0.1 dex in our models.  The changes in the mass
     fraction of helium at the surface are very small, just over 1\%
     at maximum for the tightest system. An element that is a very
     sensitive tracer of rotational mixing is boron.  It is easily
     destroyed in the hotter layers just below the surface. Therefore
     the time delay for boron, after which the boron surface abundance
     starts to change, is very short.  The change in the boron surface
     abundance is considerable, up to 1.2 dex in the 1.4 day model.
     In practice it may be hard to measure boron due to its low overall abundance.

     The closer the stars are to filling their Roche lobe, the higher
     is their surface N abundance. This can be turned around and may
     be used to predict which binary systems are likely to show
     nitrogen surface enhancements.  In Tab.~\ref{tab:smc} we indicate
     the Roche-lobe filling factor $R/R_{\rm RL}$, defined as the
     radius of the primary star over the Roche-lobe radius, at the
     moment when the surface N enhancement exceeds the initial
     abundance by 0.2 dex. Based on these models, we predict that the
     observed systems with masses close to 20 and 15\Msun~have to fill
     their Roche lobes by about 86\% or more before they show
     detectable surface N enhancements (see Table~\ref{tab:smc}
     col. 5). We do note that the radii predicted by our models are
     sensitive to the assumed amount of overshooting.

     In the last two columns of Table~\ref{tab:smc} we indicate the
     rotational velocity at the equator, averaged over time and at the
     onset of Roche-lobe overflow.  Whereas a single star at this
     metallicity slows down its rotation rate due to its evolutionary
     expansion, the spin periods of the stars in tidally locked
     binaries are nearly constant because the orbit acts as an angular
     momentum reservoir.  Therefore rotationally induced mixing is
     more important in a star in a tidally locked binary than in a
     single star that started with the same initial rotation rate.
     The effect is small, however, as the expansion mainly takes place
     after about 5--6 Myr for a 20\Msun~star.  Binaries with orbital
     periods shorter than 2 days fill their Roche lobe before the
     expansion sets in.  The effect is stronger at higher metallicity,
     where angular momentum loss by the stellar wind becomes
     important, see Sect.~\ref{sec:dis}.

     \subsubsection*{Expected trends with system mass and mass ratio}
     For an ideal comparison between models and observations one would
     prefer a model in which the binary parameters match the observed
     parameters. In the above, we discussed models with a specific
     choice for the primary mass and mass ratio and we only varied the
     initial orbital period.  Here we discuss how our findings are
     expected to change for systems with slightly different system
     masses and mass ratios.

     In more massive binaries rotational mixing is more efficient,
     such that they will show a shorter time delay, even when measured
     in units of the core hydrogen-burning lifetime.  Detectable
     surface enhancements are expected to occur at an earlier stage,
     while the stars are further away from filling their Roche lobe.
     At very high masses additional effects can play a role, which are
     discussed in Sect.~\ref{bin:lmc}.  The secondary star in the
     models discussed above does not show significant surface
     abundance changes for C, N and He. The surface boron abundance
     does decrease by up to $0.6$~dex for the 1.4 day model.  In
     systems with mass ratios closer to~1, the time scale for
     rotational mixing becomes similar in both stars and nitrogen
     enhancements are expected for both stars.

     Another effect of changing the mass ratio is that the size of the
     Roche lobe changes. A more extreme mass ratio results in a wider
     Roche lobe for the primary, if the primary mass and orbital
     period are kept constant.  The efficiency of rotational mixing
     does not change, as it depends on the rotational period which is
     fixed by the orbital period.  However, the star has more space to
     expand before it fills its Roche lobe.  We noted above that the
     primary stars fill their Roche lobes by at least 86\% before they
     show detectable surface N enhancements (see also
     Sect~\ref{sec:dis}).  This changes to approximately 83\% and 76\%
     for mass ratios of $q = 0.5$ and $q = 0.25$ respectively.  A
     bigger Roche lobe also leaves more time to enhance the surface
     abundance. This implies a greater chance of catching the system in
     a stage where the surface abundance is significantly enhanced. 
   
     To summarize: the biggest surface N enhancements are expected for
     systems with a high-mass primary, which is close to filling its
     Roche lobe, with a companion which has a significantly lower mass
     in an orbit of no more than a few days.

\begin{figure}
  \resizebox{\hsize}{!}{\includegraphics{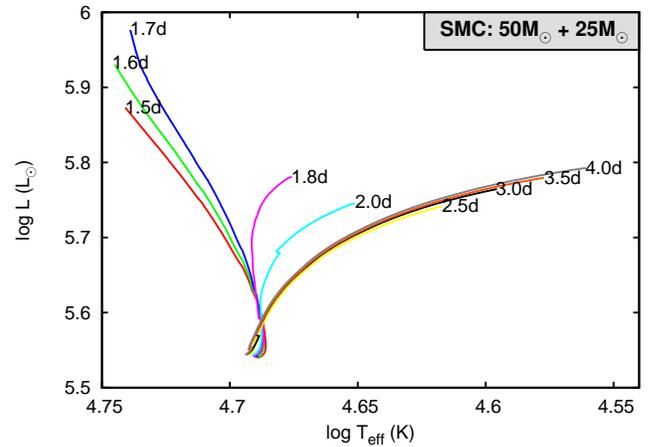}}
  \caption{The evolution from the onset of central H burning until the
    moment of Roche-lobe overflow for a 50\Msun~star in a binary
    with a 25\Msun~companion (not plotted) with initial orbital
    periods between 1.5 and 4 days. }
  \label{lmc_hrd}
\end{figure}

\begin{figure}
  \resizebox{\hsize}{!}{\includegraphics{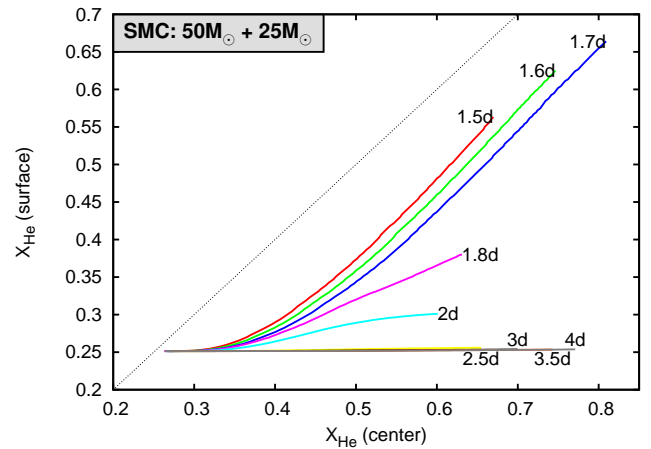}}
  \caption{ Helium abundance at the surface as a function of the helium
  abundance in the center for the same systems as plotted in
  Figure~\ref{lmc_hrd}. } \label{lmc_He}
\end{figure}

\begin{figure*}
\centering
\includegraphics[width=8.5cm]{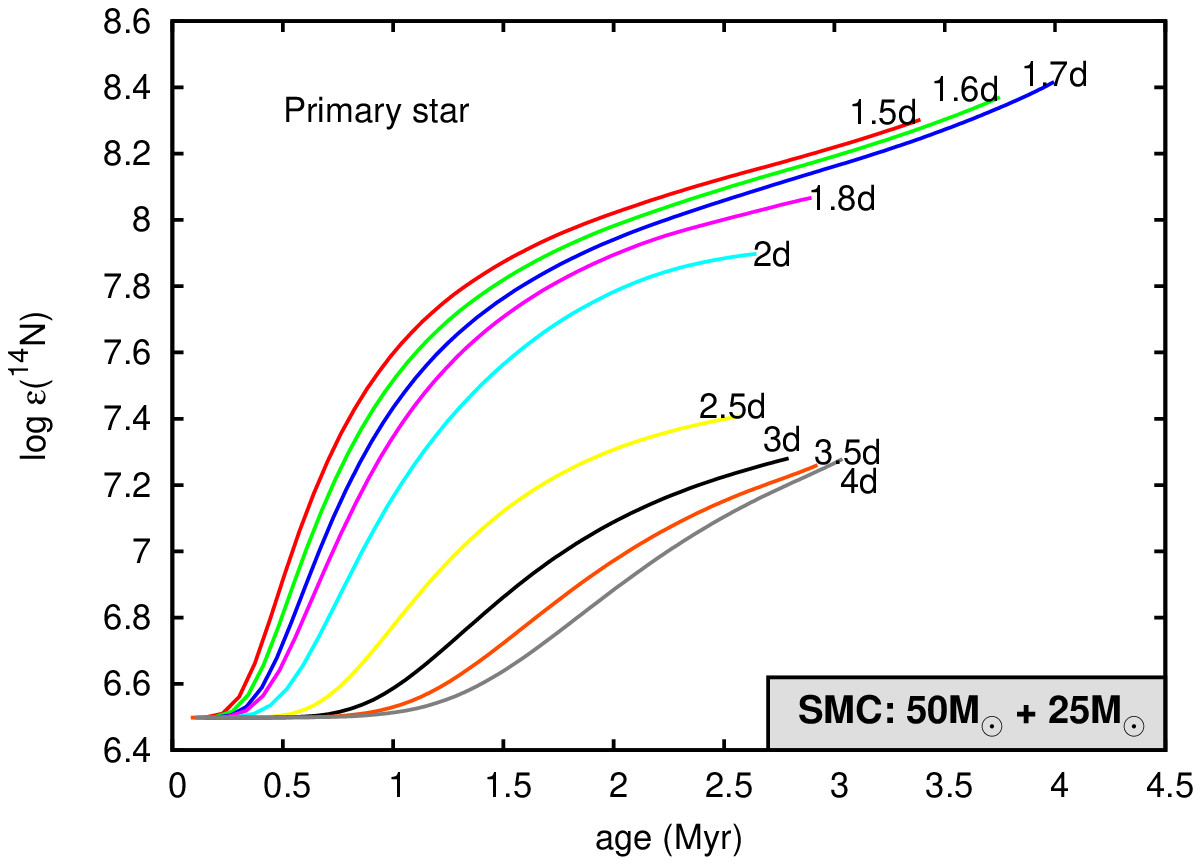}\includegraphics[width=8.5cm]{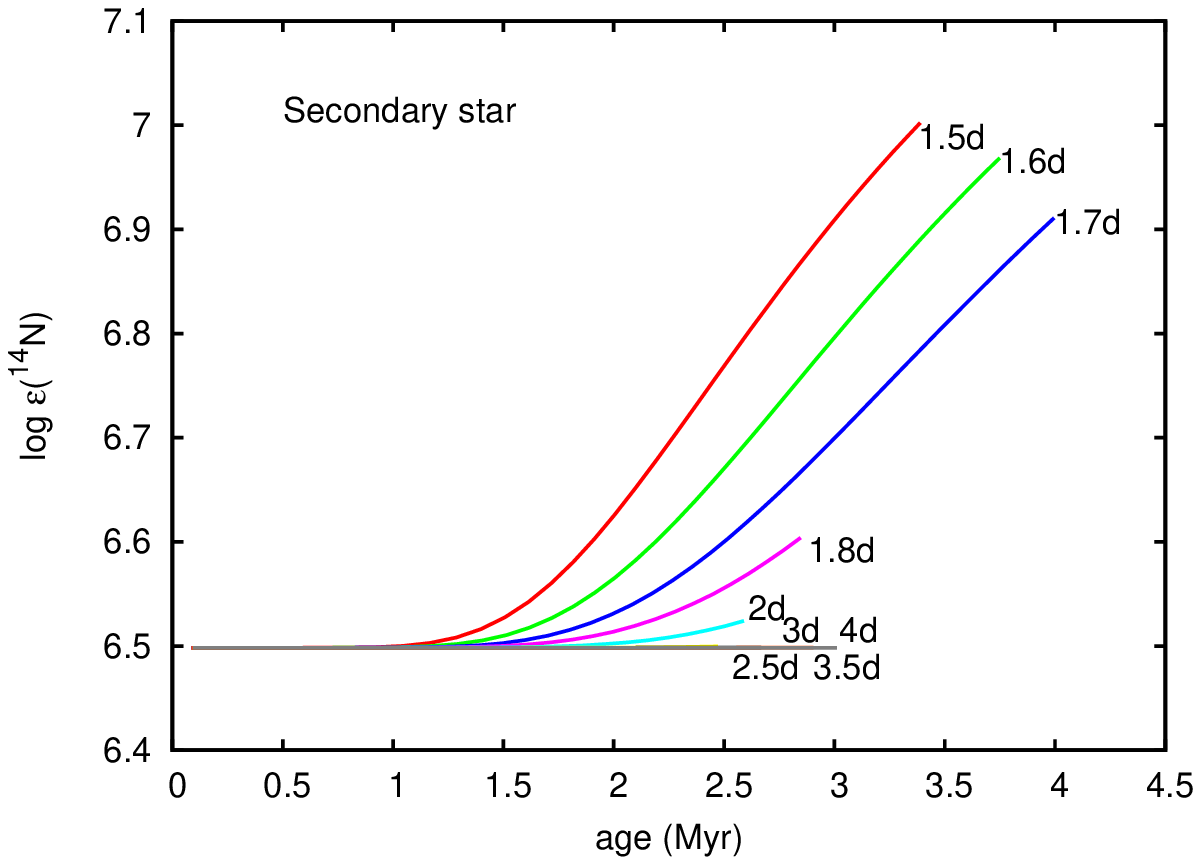}
\caption{Nitrogen abundance as a function of age at the
  surface for the primary and secondary star of the same systems as
  plotted in Figure~\ref{lmc_hrd}. Note the different scales. 
  } \label{lmc_abun}
\end{figure*}

\begin{figure*}
\centering
\includegraphics[width=8.5cm]{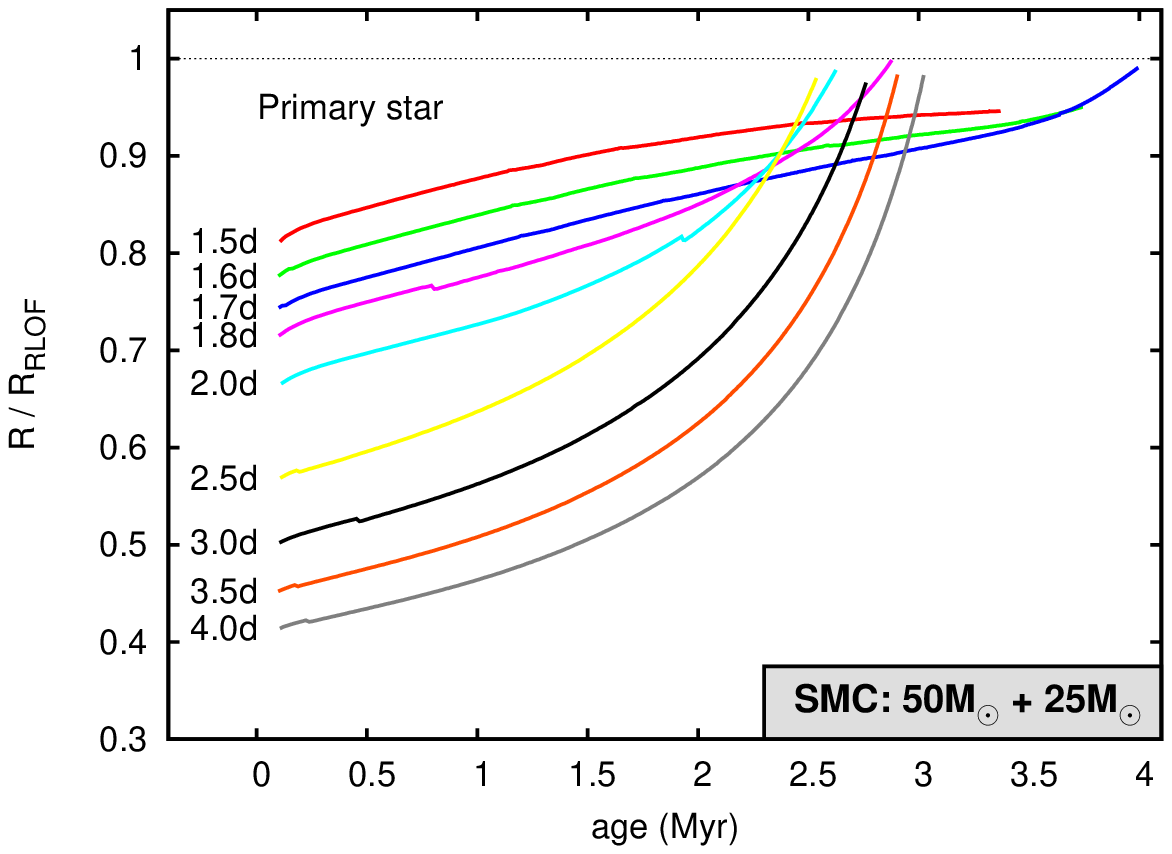}\includegraphics[width=8.5cm]{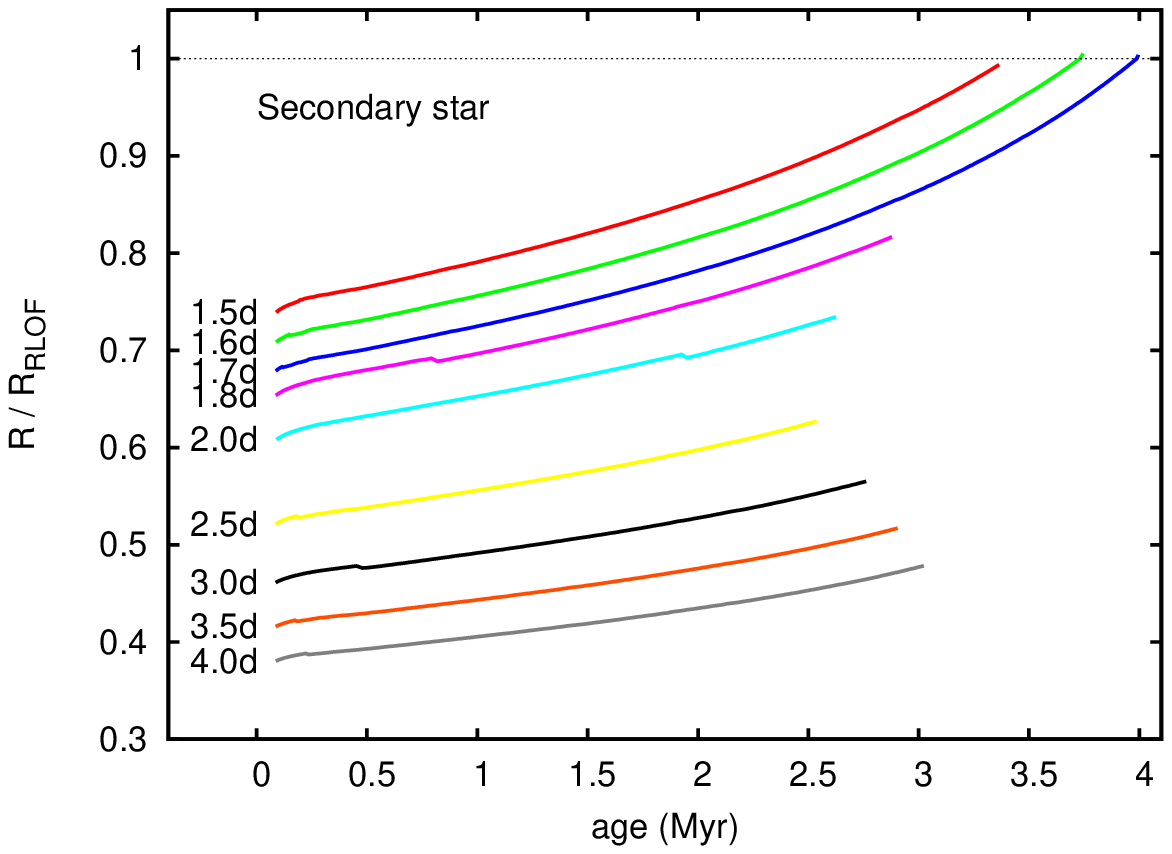}
\caption{ Radius as fraction of the Roche-lobe radius for the primary
  (left panel) and secondary star (right panel) of the same systems as
  plotted in Figure~\ref{lmc_hrd}.  } \label{lmc_rlf}
\end{figure*}

\begin{table*} 
\caption{   
    Key properties of very massive binaries (50\Msun$+$25\Msun) as
    described in Sect.~\ref{bin:lmc} (see also Table~\ref{tab:smc} and
    Sect.~\ref{bin:hil}). In the last column we indicate which
    component fills its Roche lobe first. \label{tab:lmc} }
\centering                                      
\begin{tabular}{c c c c c c c c c c c}         
\hline\hline                       
$P_{\rm orb} (d)$ & 
$t_{\rm RL}$(Myr) & 
X$_{\rm He,center}$ & 
X$_{\rm He,surface}$ &
$\log \epsilon ({\rm N}^{14})$ & 
$ {^{}{\rm B}_{\rm RL}/^{}{\rm B}_{\rm init}}$ & 
$ {^{}{\rm C}_{\rm RL}/^{}{\rm C}_{\rm init}}$ &
$ {^{}{\rm N}_{\rm RL}/^{}{\rm N}_{\rm init}}$ &
$\langle \varv_{\rm eq}\rangle$~(kms$^{-1}$) & 
$\varv_{\rm eq,RL}$~(kms$^{-1}$) &
RLOF\\ 
\hline 
   1.5 &    3.4   &    0.68 &    0.56 &    8.30 & 2.0$\times10^{-6}$ & 9.0$\times10^{-2}$ & 37  &    304 &  315 & Sec.\\
   1.6 &    3.7   &    0.75 &    0.63 &    8.37 & 6.6$\times10^{-7}$ & 9.1$\times10^{-2}$ & 37  &    289 &  307 & Sec.\\
   1.7 &    4.0   &    0.81 &    0.66 &    8.41 & 4.7$\times10^{-7}$ & 9.5$\times10^{-2}$ & 37  &    277 &  310 & Sec.\\
   1.8 &    2.9   &    0.64 &    0.38 &    8.06 & 2.8$\times10^{-4}$ & 2.3$\times10^{-1}$ & 30  &    262 &  315 & Prim.\\
   2.0 &    2.6   &    0.61 &    0.30 &    7.90 & 2.4$\times10^{-3}$ & 4.0$\times10^{-1}$ & 23  &    240 &  305 & Prim.\\
   2.5 &    2.6   &    0.66 &    0.26 &    7.41 & 8.1$\times10^{-3}$ & 7.4$\times10^{-1}$ & 8.0  &    200 &  279 & Prim.\\
   3.0 &    2.8   &    0.71 &    0.25 &    7.28 & 1.3$\times10^{-2}$ & 7.8$\times10^{-1}$ & 6.0  &    172 &  261 & Prim.\\
   3.5 &    2.9   &    0.75 &    0.25 &    7.26 & 1.8$\times10^{-2}$ & 8.0$\times10^{-1}$ & 5.7  &    152 &  248 & Prim.\\
   4.0 &    3.0   &    0.78 &    0.25 &    7.27 & 2.4$\times10^{-2}$ & 7.9$\times10^{-1}$ & 6.0  &    137 &  236 & Prim.\\

\hline 
\end{tabular}
\end{table*}

\subsection { Very massive binaries
  (50\Msun$+$25\Msun)} \label{bin:lmc} \label{sec:lmc}

In more massive binaries, rotational mixing can be so efficient that
the change in chemical profile leads to significant structural
changes. In this model set we adopt a primary mass of 50\Msun, a
secondary mass of 25\Msun~and orbital periods varying from 1.5 to 4
days, assuming an SMC composition to be consistent with the models
presented in Sect.~\ref{sec:hil}.  
Although such massive close systems are rare, observational
counterparts do exist, for example two of the four massive binaries 
presented by \citet{Massey+02} which are located in the R136 cluster
at the center of the 30 Doradus nebula in the Large Magellanic Cloud:
R136-38\footnote{$M_1=56.9\pm0.6\Msun$, $M_2=23.4\pm0.2\Msun$ and $P_{\rm
orb} = 3.39$d} and R136-42\footnote{$M_1 =40.3\pm0.1\Msun $, $M_2
=32.6\pm0.1\Msun$ and $P_{\rm orb} =2.89$d}.
 Another example with an even closer orbit is [L72]~LH~54-425\footnote{
$M_1=47\pm2\Msun$, $M_2 =28\pm1\Msun$ and $P_{\rm orb} =2.25$d}
located in the LH~54~OB association in the Large Magellanic Cloud
\citep{Williams+08}. All three binary systems have O-type main-sequence
components, which reside well within their Roche lobes.

Figure~\ref{lmc_hrd} shows the evolution of the primary stars in the
Hertzsprung-Russell diagram, until one of the stars in the binary
fills its Roche lobe  (not necessarily the primary, see
below). At the onset of hydrogen burning their location in the
diagram is very similar, although the stars in tighter binaries, which
rotate faster, are slightly cooler and bigger: a direct consequence of
the centrifugal force.  As they evolve their tracks start to deviate.
The wider systems ($P_{\rm orb} > 2.0$d) evolve similarly to
non-rotating stars: they expand during core hydrogen burning, evolving
towards cooler temperatures until they fill their Roche lobe. Their
evolutionary tracks overlap in the Hertzsprung-Russell diagram.

The primaries in the tighter systems ($P_{\rm orb} < 2.0$d) behave very
differently: they evolve left- and up-ward in the HR-diagram, becoming
hotter and more luminous while they stay relatively compact.  The
transition in the morphology of the tracks around $P_{\rm orb} \approx
2$ days is similar to the bifurcation found for fast rotating single
stars \citep{Maeder87,Yoon+Langer05}.

\subsubsection*{Surface abundances}

Rotational mixing is so efficient in these systems that even a large
amount of helium can be transported to the
surface. Figure~\ref{lmc_He} depicts the helium mass fraction at the
surface as a function of the helium mass fraction in the center.  In
the hypothetical case that mixing would be extremely efficient
throughout the whole star, the surface helium abundance would be equal
to the central helium abundance at all time. This is indicated by the
dotted line.  For the widest systems, the surface helium mass fraction
is not affected by rotational mixing at all, while for the tighter
systems $X_{\rm He}$ reaches up to 65\%. They follow the evolution of
chemically homogeneous stars.  Figure~\ref{lmc_He} also shows that for
each system, mass transfer starts before all hydrogen is converted
into helium in the center. Note that the highest central He mass
fraction when mass transfer starts is reached in the 1.7~day system
(more than 80\%), whereas on the basis of standard binary evolution
theory
\citep[e.g.][]{Kippenhahn+Weigert67} one would expect this to occur in
the widest system.  This anomalous behavior is connected to the
evolution of the radius, as discussed below.

All systems show big enhancements of nitrogen at the surface of the
primary, see Fig.~\ref{lmc_abun}. The wider systems are enhanced by up
to 0.9 dex.  In the tight systems the enhancement reaches almost 2
dex. This extreme increase is partly due to the fact that abundance is
measured relative to hydrogen, which is significantly depleted at the
surface of the primaries in the tightest systems
(cf. Table.~\ref{tab:lmc}). Also the secondary stars show nitrogen
surface enhancements, of up to 0.5 dex.

\subsubsection*{Evolution of the radius}

The increase of helium in the envelopes of the primary stars in the
tightest binaries leads to a decrease of the opacity and an increase
in mean molecular weight in the outer layers, resulting in more
luminous and more compact stars.  In Figure~\ref{lmc_rlf} we plot the
stellar radii as a fraction of their Roche-lobe radii.  The primary
stars in the wider systems expand and fill their Roche lobe after
about 2.5-3 Myr.  Contrary to what one might expect, we find that
Roche-lobe overflow is delayed in tighter binaries.  Whereas classical
binary evolution theory predicts that the primary star is the first to
fill its Roche lobe, we find instead that for systems with $P_{\rm
orb} \le 1.7$~days it is the less massive secondary star that
starts to transfer mass towards the primary. During this phase of
reverse mass transfer from the less massive to the more massive star,
the orbit widens. Nevertheless we find that, if we continue our
calculations, the two stars come into contact shortly after the onset
of mass transfer and the stars are likely to merge.

\subsubsection*{Expected trends with system mass and mass ratio}

In more massive systems the effect of rotational mixing becomes
stronger in both components. This may allow for chemically
homogeneous evolution to occur in binary systems with wider orbits.
If the mass ratio is closer to one, $M_2 \approx M_1$, the effect of
rotational mixing becomes comparable in both stars, and they may both
evolve along an almost chemically homogeneous evolution track. In that
case both stars can stay within their Roche lobe and gradually become
two compact WR stars in a tight orbit.
In a system with a more extreme mass ratio, $M_2\ll M_1$, the
secondary star will hardly evolve or expand during the core
hydrogen-burning lifetime of the primary.  Also in this case
Roche-lobe overflow may be avoided during the core hydrogen-burning
lifetime of the primary component, leading to the formation of a
Wolf-Rayet star with a main-sequence companion in a tight orbit.
The parameter space in which this type of evolution can occur will be
 examined in a subsequent paper.

\section{Discussion} \label{sec:dis} 
We have shown that rotational mixing can have important effects in
close massive binaries.  The mixing parameters in our code have been
calibrated against observations of rotating massive stars in the
VLT-FLAMES survey under the assumption that rotational mixing is the
main process responsible for the observed surface N enhancements (see
Sect.~\ref{sec:code}).  The predictions we present for close binaries
can be used to test the validity of this assumption.

In Sect.~\ref{sec:sin} we conclude that eclipsing binaries in the
Magellanic Clouds are the most promising test cases for rotational
mixing.  In these systems we expect the biggest enhancements of N at
the surface. Also, the uncertainty in the mass-loss rates plays a
less important role, as radiatively driven winds are reduced at low
metallicity.  A disadvantage of Magellanic Cloud binaries with respect
to Galactic systems is the greater distance. Long integration times may
be needed to obtain spectra with high enough quality to determine the
stellar parameters and abundances.  The advantage of using eclipsing
binaries is that we can compare directly to evolution models with
corresponding masses and orbital periods. Therefore, just a few
well-studied systems may be enough to put constraints on the
efficiency of rotational mixing.  However, ideally one would prefer a
large sample to enable a statistical comparison.

The main parameter in our code that affects the efficiency of mixing
due to rotational instabilities is \fc~(see Sect~\ref{sec:code}) which
has been calibrated directly against the surface N abundances in the
VLT-FLAMES survey.  However, the calibration involves multiple
parameters, such as \fmu~which is a measure of how effectively a
gradient in mean molecular weight can inhibit rotational mixing.  We
expect that uncertainties in this parameter are not important for
our predictions for nitrogen, as this element is mainly produced and
transported to the envelope early in the evolution, before a strong
mean molecular-weight gradient has been established at the interface
between the core and the envelope. However, a lower value of
\fmu~may lead to higher helium surface abundances, which may
facilitate the possibility of chemically homogeneous evolution in
close binaries, but this remains to be investigated.

We use a large amount of overshooting in our models (0.355 times the
pressure scale height $H_p$), to reproduce the extension of the main
sequence observed in the VLT-FLAMES data. However, the amount of
overshooting is an uncertain parameter and some authors quote lower
values for the amount of overshooting, for example
\citet{Schroeder+97} who find 0.24-0.32$H_p$ based on eclipsing
binaries with stellar masses between 2.5 and 6.5\Msun, \citep[see
also][]{Stothers+Chin92,Alongi+93}. We recomputed one of our models
(20+15\Msun, 3 days) without overshooting we find that, although the
stars are less luminous as expected, the surface nitrogen abundance
and the radius at a given time are very similar (deviations of less
0.01 dex in the N abundance at a given age, and less than 4\% in the
radius for a given nitrogen surface abundance). For tighter binaries
we expect an even smaller effect on the radius. We conclude that our
predictions are not very sensitive to the uncertainties in the
overshooting parameter.

In our models we assume that mixing processes in binaries operate in
the same way as in single stars.  For both single stars and binary
  members we find that the stellar interior rotates nearly rigidly (at
  least during the early phase of evolution of interest here) as a
  result of efficient internal angular momentum transport by magnetic
  torques. Having very similar internal rotational profiles, the only
difference in our models arises from the evolution of the rotation
rate, which in single stars is governed by angular momentum loss and
evolutionary expansion, while in binaries the tides play a major
role. However, in the binary models we consider the stars are close to
filling their Roche lobe.  They are slightly deformed in a lob-sided
way, no longer being symmetric around the rotation axis. In addition,
one side of the star is irradiated by the companion and may be
heated. Although the system is synchronized, the tides continue to
extract or deposit angular momentum from or onto the stars.  How
  such effects, induced by the presence of the companion star,
  interact with the different rotational instabilities is not well
  understood and poses an additional uncertainty on our predictions.
  If such effects are important and lead to additional mixing, our
  predictions for the N surface abundance can still be used as a test
  for rotational mixing if they are considered as lower limits to the
  expected N abundance.

\subsection*{Avoiding mass transfer in short-period binaries}

We have shown that rotational mixing, if it is as efficient as assumed
in our models, can lead to chemically homogeneous evolution for
tight binaries with a 50\Msun~primary. In these models the primary
star stays so compact that the secondary star is the first to fill its
Roche lobe.

This peculiar behavior of the radius of stars, which are efficiently
mixed, has been noted in models of rapidly rotating massive single
stars \citep{Maeder87} and has been suggested as an evolutionary
channel for the progenitors of long gamma-ray bursts
\citep{Yoon+06, Woosley+06} in the collapsar scenario
\cite{Woosley93}.  In single stars this type of evolution only occurs
at low metallicity, because at solar metallicity mass and angular
momentum loss in the form of a stellar wind spins down the stars and
prevents initially rapidly rotating stars from evolving chemically
homogeneously \citep{Yoon+06, Brott+09}. In a close binary tides can
replenish the angular momentum, opening the possibility for chemically
homogeneous evolution in the solar neighborhood.

The binary models presented here all evolve into contact, but (as we
discussed briefly in Sect~\ref{sec:lmc}) Roche-lobe overflow may be
avoided altogether in systems in which the secondary stays compact,
either because it also evolves chemically homogeneously, which may
occur if $M_1 \approx M_2$, or because it evolves on a much longer
timescale than the primary, when $M_2\ll M_1$. Whereas standard binary
evolution theory predicts that the shorter the orbital period, the
earlier mass transfer sets in, we find that binaries with the lowest
orbital periods may avoid the onset of mass transfer altogether.  This
evolution scenario does not fit in the traditional classification of
interacting binaries into Case~{\it A}, {\it B} and {\it C}, based on
the evolutionary stage of the primary component at the onset of mass
transfer \citep{Kippenhahn+Weigert67,Lauterborn70}.  In the remainder
of this paper we will refer to this new case of binary evolution, in
which mass transfer is delayed or avoided altogether as a result of
very efficient internal mixing, as Case~{\it M}.

The massive and tight systems in which Case~{\it M} can occur are
rare. Additional mixing processes induced by the presence of the
companion star, which may be important in such systems, will widen the
parameter space in which Case~{\it M} can occur: it would lower the minimum
mass for the primary star and increase the orbital period below which
this type of evolution occurs.  The massive LMC binary
[L72]~LH~54-425, with an orbital period of 2.25~d
\citep[][see also Sec.~\ref{sec:lmc}]{Williams+08} may be a candidate
for this type of evolution. Another interesting case is the galactic
binary WR20a, which consists of two core hydrogen burning stars of
$82.7\pm5.5$ and $81.9\pm5.5\Msun$ in an orbit of 3.69~d. Both stars
are so compact that they are detached. The surface abundance show
evidence of rotational mixing: a nitrogen abundance of six times
solar is observed and carbon is depleted \citep{Bonanos+04, Rauw+05}.

\subsection*{Short-period Wolf-Rayet and black-hole binaries  }

If Roche-lobe overflow is avoided throughout the core hydrogen-burning
phase of the primary star, both stars will stay compact while the
primary gradually becomes a helium star and can  be observed as a
Wolf-Rayet star.  Initially the Wolf-Rayet star will be more massive
than its main sequence companion, but mass loss due to the strong
stellar wind may reverse the mass ratio, especially in systems which
started with nearly equal masses. Examples of observed short-period
Wolf-Rayet binaries with a main-sequence companion are
CQ~Cep\footnote{CQ~Cep: ${\rm M_{WR}}= 24\Msun,\,{\rm
M_O}=30\Msun,\,P_{\rm orb} = 1.6$~d}, CX~Cep\footnote{CX~Cep: ${\rm
M_{WR}}=20\Msun,\,{\rm M_O}=28\Msun,\, P_{\rm orb} = 2.1$~d},
HD~193576\footnote{HD~193576: ${\rm M_{WR}}=9\Msun,\,{\rm
M_O}=29\Msun,\, P_{\rm orb} = 4.2$~d} and the very massive system
HD~311884\footnote{HD~311884: ${\rm M_{WR}}=51\Msun,\,{\rm
M_O}=60\Msun,\, P_{\rm orb} = 6.2$~d} \citep{vanderHucht01}. Such
systems are thought to be the result of very non-conservative mass
transfer or a common envelope phase
\citep[e.g.][]{Petrovic+05_WR}. Case~M constitutes an alternative
formation scenario which does not involve mass transfer.

Case~{\it M} is also interesting in the light of massive black-hole
binaries. \citet{Orosz+07} recently published the stellar parameters
of M33~X-7, located in the nearby galaxy Messier~33 which harbors one
of the most massive stellar black holes known to date, ${\rm
M_{bh}}=15.7\pm1.5\Msun$, orbiting a massive O star, ${\rm M_O}=70\pm
7\Msun$, which resides inside its Roche lobe in spite of the fact that
the orbit is very tight, $P_{\rm orb} = 3.45$~d.  The explanation for
the formation of this system with standard binary evolutionary models
involves a common-envelope phase that sets in after the end of core
helium burning (Case~{\it C}), as the progenitor of the black hole
must have had a radius much greater than the current orbital
separation.  This scenario is problematic as it requires that the
black-hole progenitor lost roughly ten times less mass before the
onset of Roche-lobe overflow than what is currently predicted by
stellar evolution models
\citep{Orosz+07}.  An additional problem is that the most likely
outcome of the common envelope phase would be a merger, as the
envelopes of massive stars are tightly bound
\citep{Podsiadlowski+03}. In the Case~{\it M} scenario the black-hole
progenitor can stay compact and avoid Roche-lobe overflow at least
until the end of core helium burning, such that it retains its
envelope. 

There are examples of stellar mass black-hole binaries with short
periods and a massive main-sequence companion, in which nearly
chemically homogeneous evolution may be important. IC~10~X-1 is a 
system harboring the most massive stellar mass black hole known to
date with a mass of at least 21\Msun, orbiting a Wolf-Rayet star of
approximately 25\Msun~in an orbit of 1.45 days
\citep{Silverman+Filippenko08}. Homogeneous evolution helps to explain
the high mass of the black hole, but the short orbital period poses a
difficulty, also for Case~{\it M}: strong mass loss during the Wolf-Rayet
life time will widen the orbit. Other examples of high mass black hole
binaries with short orbital periods are the famous systems
Cyg~X-1\footnote{Cyg~X-1: ${\rm M_{bh}}\approx 10\Msun,\,{\rm
    M_O}\approx18\Msun,\,P_{\rm orb}=5.6$~d} \citep{Herrero+95},
LMC~X-1\footnote{LMC~X-1: ${\rm M_{bh}} \approx 10\Msun,\,{\rm
    M_O}\approx30\Msun,\, P_{\rm orb}=3.9$~d} \citep{Orosz+08} and
LMC~X-3\footnote{LMC~X-3: ${\rm M_{bh}}\approx4$--$10\Msun,\,{\rm
    M_O}\approx40\Msun,\,P_{\rm orb}=4.2$~ d} \citep[and references
therein]{Yao+05}.

The subsequent evolution of tight rapidly rotating Wolf-Rayet binaries
remains to be investigated. If one or both members of the 
system can retain enough angular momentum to fulfill the collapsar
scenario \citep{Woosley93}, which may be hard as the tides can slow
down the stars \citep[e.g.][]{Detmers+08}, it may lead to the
production of one or even two long gamma-ray bursts. 

Although promising, the importance of this channel may be small due to
the limited binary parameter space in which Case~{\it M} can occur.  To make
any strong statements about this new evolutionary scenario, further
modeling is needed, which we will undertake in the near future.

\section{Conclusion} \label{sec:con}  

We investigated the effect of rotational mixing on the evolution of
detached short-period massive binaries using a state of the art
stellar evolution code.  The efficiency of rotational mixing was
calibrated under the assumption that rotational mixing is the main
process responsible for the observed N enhancements in rotating stars.

We find nitrogen surface enhancements of up to 0.6 dex for massive
binaries in the SMC.  The largest enhancements can be reached in
systems with orbital periods less than about 2 days, in which the
primary is massive (about 20\Msun~or more) and evolved (filling its
Roche lobe by about 80\% or more) and the secondary is significantly
less massive, which leads to a more spacious Roche lobe for the
primary (preferably $M_2/M_1 \lesssim 0.75$). 

We propose to use such systems as test cases for rotational mixing.
These systems often show eclipses and big radial velocity
variations, such that their stellar parameters, the rotation rate and
possibly their surface abundances can be determined with high
accuracy.  This enables a direct comparison between an observed system
and models computed with the appropriate stellar and binary
parameters.  An additional major advantage of using detached
main-sequence binaries is the constraint on the evolutionary
history. For a fast spinning apparently single star we do not know
whether it was born as a fast rotator or whether its rotation rate is
the result of mass transfer or merger event. In a detached
main-sequence binary we can exclude the occurrence of any mass transfer
phase since the onset of core-hydrogen burning (see
Sect~\ref{sec:intro}).

In the most massive binaries we find that rotational instabilities can
efficiently mix centrally produced helium throughout the stellar
envelope of the primary. They follow the evolution for chemically
homogeneous stars: they stay within their Roche lobe, being
over-luminous and blue compared to normal stars. Due to large amount
of nitrogen and helium at the surface these stars can be observed as
Wolf-Rayet stars with hydrogen in their spectra. In contrast to
standard binary evolution, we find that it is the less massive star in
these systems that fills its Roche lobe first.

There may be regions in the binary parameter space in which Roche-lobe
overflow can be avoided completely during the core hydrogen-burning
phase of the primary. The parameter space for this new evolutionary
scheme, which we denote Case~{\it M} to emphasize the important role of
mixing, increases if additional mixing processes play a role in such
massive systems.  It may provide an alternative channel for the
formation or tight Wolf-Rayet binaries with a main-sequence companion,
without the need for a mass transfer and common envelope phase to
bring the stars close together.  This scenario is also potentially
interesting for tight massive black hole binaries, such as M33~X-7
\citep{Orosz+07}, for which no satisfactory evolutionary scenario
exists to date.

\begin{acknowledgements}
  The authors would like to thank G.~Meynet (referee), E.~Glebbeek,
  M,~Verkoulen, K.~Belzcinski, Z.~Han, J.-P.~Zahn and the members of
  the VLT-FLAMES consortium.  The authors acknowledge NOVA and LKBF for
  financial support. S.-Ch. Y. is supported by the DOE SciDAC Program
  (DOE DE-FC02-06ER41438).
\end{acknowledgements}

\bibliographystyle{aa}
\bibliography{references}

\end{document}